\documentclass[review]{elsarticle}

\usepackage{lineno,hyperref}
\usepackage{amsfonts}
\usepackage{epstopdf}
\usepackage[export]{adjustbox}
\usepackage{float}
\usepackage{graphicx}
\usepackage{subfigure}
\usepackage{color}
\usepackage{bm}
\usepackage{textcomp} 
\usepackage{enumitem}
\hypersetup{
	colorlinks=true,
	urlcolor= blue,
	citecolor=blue,
	linkcolor= blue,
	bookmarksopen=false,
}
\modulolinenumbers[5]

\journal{journal of Physica E}

\begin{document}

\begin{frontmatter}

\title{Bias-Voltage-Induced Topological Phase Transition in Finite Size Quantum Spin Hall Systems in the Presence of a Transverse Electric Field}

\author{A. Baradaran}
\ead{a.baradaran@ut.ac.ir}
\author{M. Ghaffarian}
\address{Department of Physics, University of Qom, Qom, Iran}

\begin{abstract}
Using the tight-binding BHZ model and Landauer-Büttiker formalism, the topological invariant of the finite width of ribbons of HgTe/CdTe quantum well is studied in the absence and presence of an external transverse electric field. It will be recognized that a critical current changes topological invariant of ribbons of quantum well. This topological phase transition, which occurred by adjustment of the bias voltage, depends on the width of the sample and the gate voltage. The profound effects of an external transverse electric field are considered to the separation of spin-up and spin-down band structures, decreasing band gap and tuning the topological phase transition between ordinary and quantum spin Hall regime. These declares the transverse electric field amplifies the quantum spin Hall regime and causes inducing the topological phase transition in ribbons of quantum well. Our finding may instantly clear some practical aspects of the study in the field of spintronic for employment in spin-based devices.
\end{abstract}


\end{frontmatter}


\section{\label{sec:Intro}Introduction}

The discovery of integer and fractional quantum Hall effect (QHE) on the grounds of experiment and theory was the first encounter with topological quantum state of matters~\cite{klitzing_PhysRevLett_45_1980_QHE,laughlin_PhysRevB.23.1981_QHC,laughlin_PhysRevLett.50.1983_AGHE,thouless_physc.SolidStatePhys.14.1981_2DHE,tsui_PhysRevLett.48.1982_2DMagR}. The features of this discovery, which is on the threshold of topological state, were very strange that scientists could not fit with the paradigms of ordinary condensed matter physics involving the trivial topological state. In two-dimensional integer quantum Hall effect (IQHE), quantum calculations show that Landau levels, in the presence of a uniform magnetic field, with a distinct gapped between HOMO and LUMO are analogous with an intrinsic semiconductor. In addition, from a straightforward calculation it was deduced that the universal value for Hall conductance is $G_{xy}=\frac{ne^2}{h}$, where $n$ represents an integer known as topological invariant or TKNN invariant~\cite{thouless_PhysRevLett.49.1982_QHCPP}. The Berry phase formula based on his well-known geometric phase of adiabatic quantum mechanics~\cite{berry_RoyalScociety_1802.1984_BerryPh}, and the TKNN formula were connected to shows that they would play an important role in classifying quantum states based on “Chern number” or “first Chern class”, given by $\frac{1}{2\pi}$ times the integral of a Berry curvature over a 2D manifold. It was recognized that necessary condition for QHE was not the existence of a magnetic field. Breaking time-reversal symmetry by applying a simple model, named Haldane model~\cite{haldane_PhysRevLett.61.1988_QHEWOLL}, showed that there are quantum Hall conductance and edge current without any Landau levels, so topological state can exist in the absence of an external magnetic field.

The idea of quantum spin Hall effect (QSHE) was derived from the Haldane model by C.L.Kane and E.J.Mele in 2005~\cite{kane_PhysRevLett.95.2005_QSHE,kane_PhysRevLett.95.2005_Z2QSHE}. This was an initial proposal~\cite{kane_PhysRevLett.95.2005_QSHE} of a topological insulator in graphene. They combined two conjugate copies of the Haldane model, one for spin-up electrons for which the valence band had Chern number $\pm 1$ and one for spin-down electrons where the valence band had the opposite value $\mp 1$; on the edges, spin-up and spin-down edge modes propagated in opposite directions. The numerical calculation showed that, as long as time-reversal symmetry was not broken, the edge modes were protected by a $\mathbb{Z}_2$ topological invariant linked to Kramers degeneracy~\cite{fu_PhysRevB.76.2007_TIIS}. Therefore, the edges consisted of counter-propagating states with opposite spin-polarization on each edge~\cite{skolasinski_PhysRevB.98.2018_HESQSHR}. Indeed, in QSHE, the Hall conductance is zero, $G_{xy}=G_{xy}^{\uparrow}+G_{xy}^{\downarrow} = 0$, but the spin Hall conductance is non-zero, $G_{xy}^{(s)}=G_{xy}^{\uparrow}-G_{xy}^{\downarrow} \neq 0$.

However QSHE, because of week spin-orbit effect of carbon, have not been observed in graphene but in 2006, Bernevig, Hughes and Zhang using k.p theory calculate the band structures of HgTe/CdTe quantum wells and showed that the energy bands near the Fermi energy have topological insulator properties. They, in their model, derived an effective four-band Hamiltonian near the Fermi energy that shows QSHE characteristics, named BHZ model and predicted QSHE are observable
experimentally due to strong spin-orbit effect of elements of quantum well. In 2007, König and their colleagues showed the edge channels could be observed in the ballistic regime~\cite{konig_PhysSocietyofJapan_77.2008_QSHE}, and these confirmed also by Roth et al.~\cite{Roth_APS2009}. Next, the researches into two-dimensional QSHE have extended for room-temperature realistic materials as a growing field~\cite{xu_PhysRevB.2015,Wang_APL.2017,sheng_ACS.2017}, especially using the BHZ model~\cite{skolasinski_PhysRevB.98.2018_HESQSHR,michetti_PhysRevB2012}. Some authors also focused on finite-size effects on edge states of quantum wells in the QSHE materials~\cite{zhou_PhysRevLett.101.2008_FSQSHE,Imura_PhysRevB.2010}. They considered the edge states and topological phase of finite-size of the quantum wells by analysis of band structures of these systems. The finite size of quantum well is quantum well that are finite in one direction and infinite in another direction, so this system are like the ribbon which we refereed as a quantum well ribbon (QWR).

In this paper, using tight-binding BHZ (TB-BHZ) Hamiltonian, the band structures of QWRs are obtained and QSHE are studied by separation of the spin-up and spin-down electrons as two distinct spin-filtered quantum Hall systems from a transport viewpoint. The transport study of QHE was performed for explanation of the quantum Hall conductance~\cite{zhou_PhysRevLett.101.2008_FSQSHE}, but in the spin-filtered quantum Hall system, there are not any Landau levels that distinguish this system from the magnetic quantum Hall system. Therefore, this motivated us to consider the QSH system profoundly that makes some interesting results like a phase transition from topological insulator to ordinary by adjustment of the bias voltage. This phenomena will be more visible in the narrower width of QWRs. In the following, we will demonstrate that this phase transition will be tuned by applying a transverse electric field.

We will organize the rest of this paper as follows. In Sec.~\ref{sec:HM}, we will introduce tight-binding BHZ (TB-BHZ) Hamiltonian in the presence of a transverse electric field for QWRs. This Hamiltonian is highly effective for low-energy. Indeed, our results will be presented in Sec.~\ref{sec:Result} that they will be in a good agreement with results have presented by the continuum BHZ model \cite{zhou_PhysRevLett.101.2008_FSQSHE,Imura_PhysRevB.2010,liu_AppPhysLett.99.2011_EFQSHE,wada_PhysRevB.83.2011_Bi,medhi_IOP_24.2012_ESTIs,hsu_PhysRevB.95.2017_InAsGaSbMagF,krishtopenko_PhyRevB_97.2018_HgTeQW}. In spite of the results have presented in Ref.~\cite{zhou_PhysRevLett.101.2008_FSQSHE}, in our approach, we will calculate the spatial distribution of spin and charge currents in the QWRs sandwiched between two metallic leads as a function of the Fermi level position at zero external electric fields and in the presence of a transverse electric field. In addition, we will briefly discuss experimental detection of QSH state in Sec.~\ref{sec:Discuss}, and mention that QWRs can be a tunable topological insulators under an appropriate transverse electric field, gate and bias voltages for application in spintronic device, which will be discussed in this paper. Finally, we will conclude the paper with Sec.~\ref{sec:Summary} where the summary of results and conclusions will be presented.

\section{\label{sec:HM}Hamiltonian Model}

We use the effective BHZ Hamiltonian that was derived
for the topological phase of HgTe/CdTe quantum wells. The BHZ
Hamiltonian in two-dimensional is in the form of $4\times4$
matrix:

\begin{equation}
\label{eqn:1}%
H(\vec{k})=
\left[ {\begin{array}{cc}
	h(\vec{k}) & 0  \\
	0 &h^{*}(-\vec{k})\\
	\end{array} } \right],
\end{equation}

\noindent the upper block
$h(\vec{k})=\varepsilon(\vec{k})\textbf{I}_{2}+\vec{d}(\vec{k}).\vec{\sigma}$,
which is for spin-up electrons, is a $2\times2$ matrix, where
$\textbf{I}_{2}$ is the unitary matrix,
$\varepsilon(\vec{k})=-D(k_{x}^{2}+k_{y}^{2})$,
$\vec{\sigma}=(\sigma_{x},\sigma_{y},\sigma_{z})$ are the Pauli
matrices, and  $\vec{d}(\vec{k})=(d_{x},d_{y},d_{z})$ are composed
from $d_{x}=Ak_{x}$, $d_{y}=Ak_{y}$, and
$d_{z}=M-B(k_{x}^{2}+k_{y}^{2})$. The lower block,
$h^{*}(-\vec{k})$, which is also for spin-down electrons, is deduced
from $h(\vec{k})$ by applying time-reversal symmetry. A, B, D, and
M are material parameters that are varied with the thickness of
quantum wells. For HgTe/CdTe quantum wells, this parameters are
adopted from Ref \cite{konig_PhysSocietyofJapan_77.2008_QSHE}, $A=364.5~ meV nm$, $B=-686~ meV nm^{2}$, $D=-512
~meV nm^{2}$, and $M=-10~ meV$.

Since band structures near $\Gamma$ point are more important, we
could write BHZ Hamiltonian for a simplified square lattice that is called as TB-BHZ model,
implemented by
$\varepsilon(\vec{k})=-2\bar{D}(2-\cos(k_{x}a)-\cos(k_{y}a))$,
$d_{x}=\bar{A}\sin(k_{x})$, $d_{y}=\bar{A}\sin(k_{y})$, and
$d_{z}=M-2\bar{B}(2-\cos(k_{x}a)-\cos(k_{y}a))$, where
$\bar{A}=Aa^{-1}$, $\bar{B}=Ba^{-2}$, and $\bar{D}=Da^{-2}$. So explicit matrix form of $h(\vec{k})$ becomes

\begin{equation} 	\label{eqn:2}
h(\vec{k})=
\left[ {\begin{array}{cc}
	M-2\bar{B}_{+}(2-\cos(k_{x}a)-\cos(k_{y}a)) & \bar{A}\sin(k_{x}a)-i\bar{A}\sin(k_{y}a)\\
	\bar{A}\sin(k_{x}a)+i\bar{A}\sin(k_{y}a) & -M+2\bar{B}_{-}(2-\cos(k_{x}a)-\cos(k_{y}a))
	\end{array} } \right],
\end{equation}

\noindent with $\bar{B}_{\pm}=\bar{B}\pm \bar{D}$. For HgTe/CdTe quantum
well we take  $\bar{A}=364.5~ meV$, $\bar{B}=-686~ meV$, and
$\bar{D}=-512~ meV$. So in our calculations, the lattice parameter is
set $a=1~ nm$.

However, we are interested to solve BHZ Hamiltonian in the finite size, so we take the upper block of TB-BHZ Hamiltonian in real space in the form of second quantization 

\begin{equation}
\label{eqn:3}%
\hat{h}=\sum_{i,j}( \underline{\underline{\varepsilon}} C^{\dagger}_{i,j} C_{i,j}
+ \underline{\underline{t}} C^{\dagger}_{i+1,j} C_{i,j}
+\underline{\underline{\acute{t}}} C^{\dagger}_{i,j+1} C_{i,j})+h.c,
\end{equation}

\noindent where $ \underline{\underline{\varepsilon}}=
\left[ {\begin{array}{cc}
	M-4\bar{B}_{+} & 0  \\
	0 & -M+4\bar{B}_{-}
	\end{array} } \right], $ is on-site matrix, $ \underline{\underline{t}}=
\left[ {\begin{array}{cc}
	\bar{B}_{+} & -i\frac{\bar{A}}{2}  \\
	-i\frac{\bar{A}}{2} & -\bar{B}_{-}
	\end{array} } \right] $, and $ \underline{\underline{\acute{t}}}=
\left[ {\begin{array}{cc}
	\bar{B}_{+} & -\frac{\bar{A}}{2}  \\
	\frac{\bar{A}}{2} & -\bar{B}_{-}
	\end{array} } \right], $ are hopping matrix in $x-$direction and $y-$direction respectively. $ C^{\dagger}_{i,j} $ and $ C_{i,j} $ are also creation and annihilation operators in site $ \vec{R}=ia\hat{x}+ja\hat{y} $ for two bands model. By converting this Hamiltonian into k-space, by taking $C^{\dagger}_{i,j}=\sum_{\vec{k}}e^{i \vec{k}.\vec{R}}C^{\dagger}_{\vec{k}}$, and $C_{i,j}=\sum_{\vec{k}} e^{-i\vec{k}.\vec{R}}C_{\vec{k}}$, we get $ \hat{h}=\sum_{\vec{k}}h(k)C^{\dagger}_{\vec{k}}C_{\vec{k}} $, where $C^{\dagger}_{\vec{k}}$ and $C_{\vec{k}}$ are creative and annihilation operators in k-space for two-bands model, and $ h(k) $ is the Hamiltonian matrix in Eqn.~\ref{eqn:2}.

\begin{figure}[!h]
	\centering
	\includegraphics[width=0.5\textwidth]{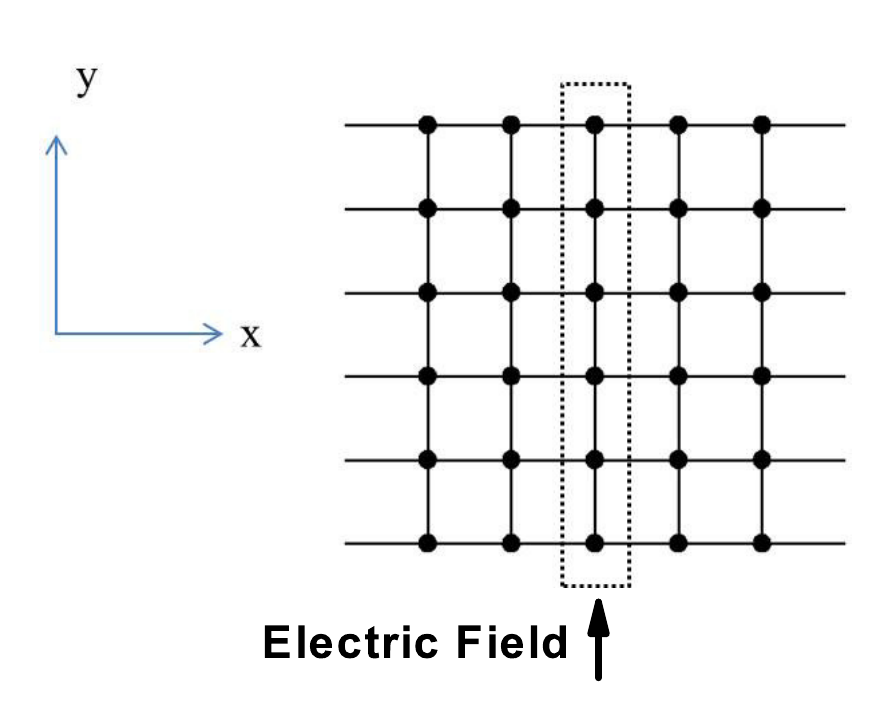}
	\caption{\label{fig:1} Schematic view of finite-size of QWR that is finite in $y-$direction, and infinite in $x-$direction. A transverse electric field may be applied in $y-$direction, also.  }
\end{figure}

For ribbons that is finite in $ y $ direction and infinite in $ x $ direction (Fig.~\ref{fig:1}), the primitive cell is chosen, and $ \hat{h} $ becomes

\begin{equation}
\label{eqn:4}%
\hat{h}=e^{-\imath ka}\hat{h}^{\dagger}_{1}+\hat{h}_{0}+e^{\imath ka}\hat{h}_{1},
\end{equation}  
\noindent where $ \hat{h}_{0}=\sum_{j}(\underline{\underline{\varepsilon}} C^{\dagger}_{i_{0},j} C_{i_{0},j}
+\underline{\underline{\acute{t}}} C^{\dagger}_{i_{0},j+1} C_{i_{0},j})+h.c $ is TB-BHZ Hamiltonian in the primitive cell, and $ \hat{h}_{1}=\sum_{j}\underline{\underline{t}} C^{\dagger}_{i_{0},j} C_{i_{0}+1,j} + h.c $ is hopping Hamiltonian between neighboring cells. Here $ i_{0} $ shows position of the primitive cell in $ x $ direction. 

By applying a transverse electric field in $y-$direction, as schematically depicted in Fig.\ref{fig:1}~ , the term $ eEy_{j} $ will have added to diagonal elements of the Hamiltonian matrix that $ e $ and $ E $ are the charge of electrons and the intensity of electric field respectively.

\section{\label{sec:Result}Results and discussion}

\begin{figure}[!h]
	\centering
	\includegraphics[width=0.8\textwidth]{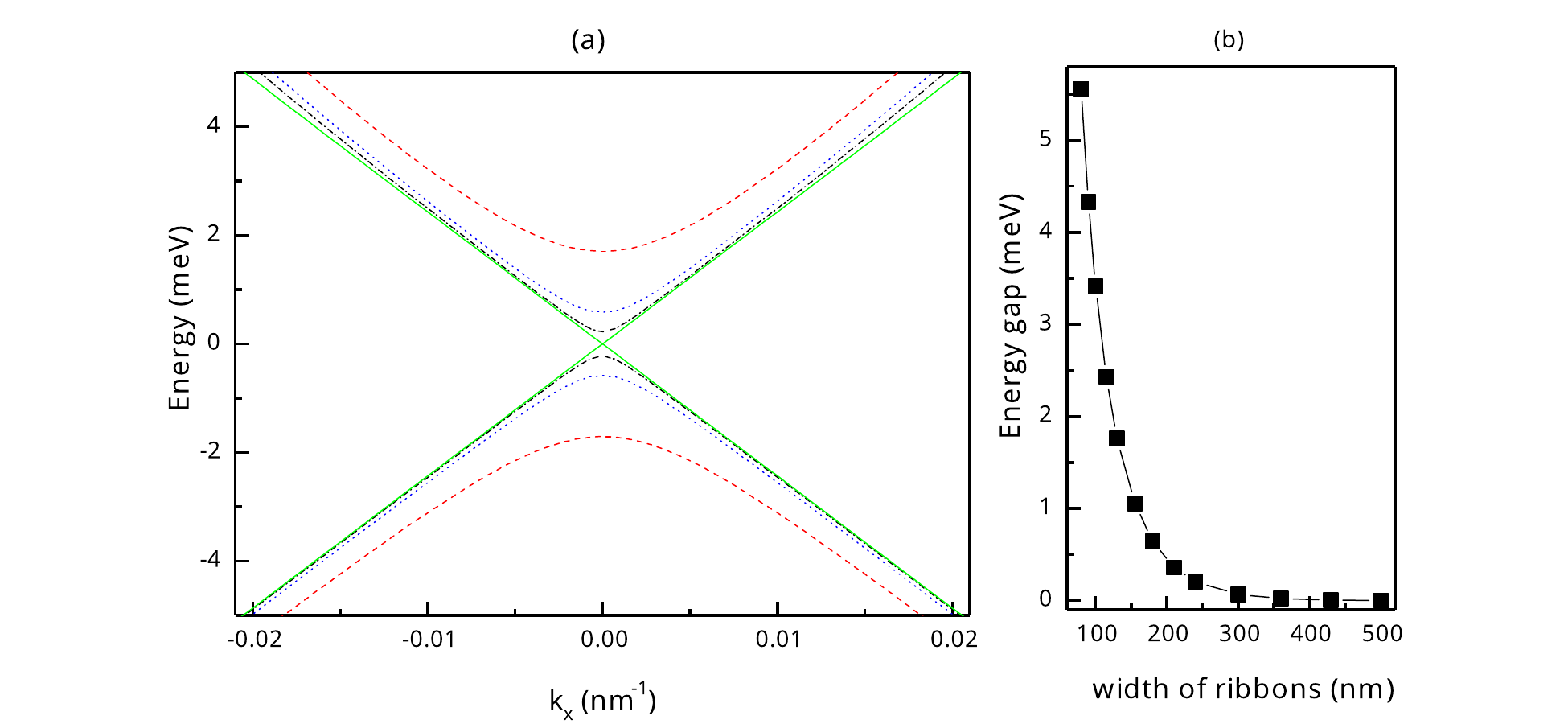}	
	\caption{\label{fig:2}(color online). (a) Band structures of QWRs with different width, 100 nm (red, dashed), 150 nm (blue, dot), 200 nm (black, dashed dot), and 500 nm (green, solid), show that energy gap closes with gradual increase of the width of QWRs, (b) Energy gap versus the width of QWRs are plotted. }
\end{figure}

Using the TB-BHZ Hamiltonian, introduced for the QWRs of HgTe/CdTe in Sec.~\ref{sec:HM}, we would calculate band structures with different width and plot two bands near Fermi energy for each QWR in Fig.~\ref{fig:2}(a). In whole computation, the Fermi level is set into zero energy. For narrower QWRs, there is a distinct gap between valence and conduction bands that is reduced dramatically by a gradual increase in the width of the QWRs, and stabilized almost after 300-nm QWR (Fig.~\ref{fig:2}(b)). These results, of course, obtained by Zhou \textit{et. al} using continuum B‌‌HZ Hamiltonian\cite{zhou_PhysRevLett.101.2008_FSQSHE}, but in the following we will show that the QSH regime could exist even in narrow QWRs in transport viewpoint. Other results from this perspective, such as variation of topological invariant as a function of gate and bias voltages will be discussed.

\vspace{-\abovedisplayskip}
\subsection{\label{sec:TW}Transport Viewpoint of QSHE in QWR}

\begin{figure}[!h]
	\centering
	\begin{tabular}[b]{cc}
		\begin{tabular}[b]{c}
			\begin{minipage}[b]{0.38\textwidth}
				\includegraphics[width=\textwidth]{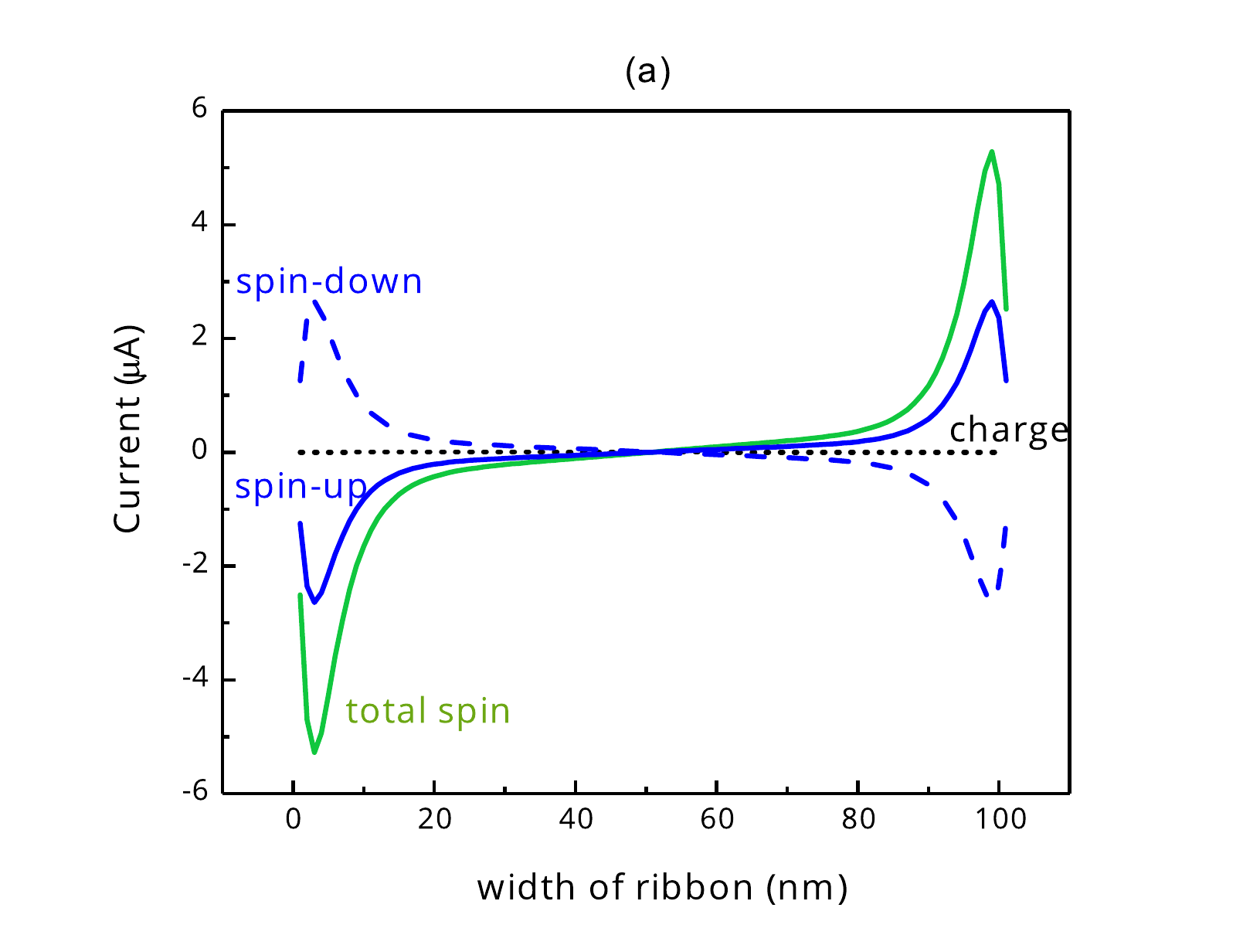}
			\end{minipage}\\
			\begin{minipage}[b]{0.29\textwidth}
				\includegraphics[width=\textwidth]{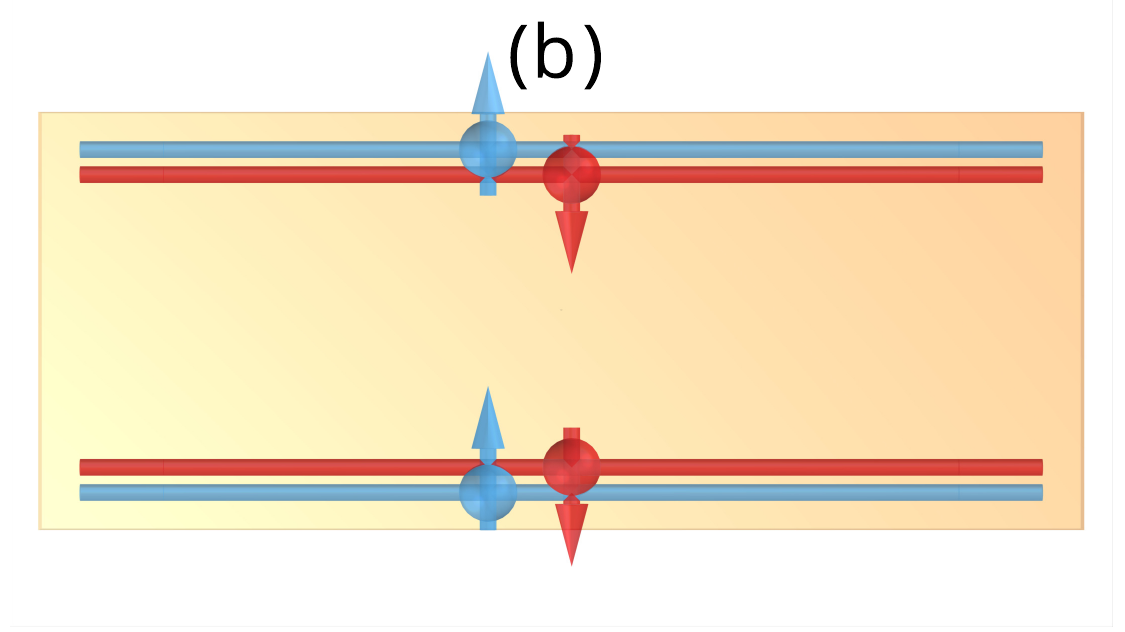}
			\end{minipage}
		\end{tabular}
		&
		\hspace{-\abovedisplayskip}
		\begin{minipage}[b]{0.6\textwidth}
			\includegraphics[width=\textwidth]{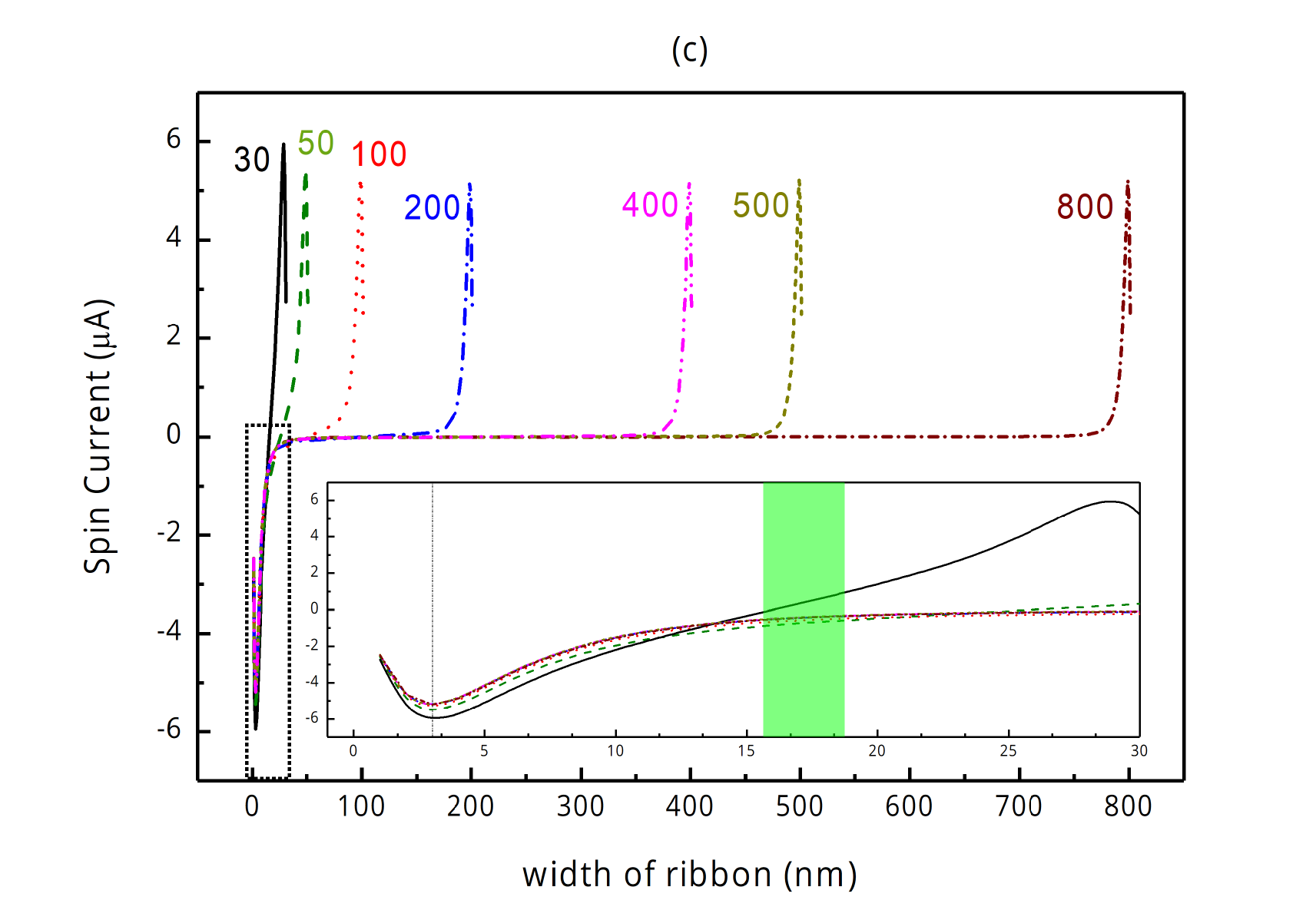}
		\end{minipage}
	\end{tabular}
	\caption{\label{fig:3}(color online). (a) Total charge current (black, dot) is remained stable, around zero, despite total spin current (green, gray) shows non-zero values along the width of QWR. Localized spin-up current (blue, solid), and localized spin-down current (dashed) match with existence a net spin current along the edge of QWR. (b) The schematic view of spin-up (blue, lighter), and spin-down (red, darker) currents along QWR is shown. (c) The LSC is depicted for different widths of QWR, 30~nm (black, solid), 50nm (olive, dashed), 100~nm (red, dot lines), 200~nm (blue, dashed dot), 400~nm (magenta, dashed dot dot), 500~nm (dark yellow, short dashed), and 800~nm (wine, short dashed dot); inset figure shows the strength of the LSC decreases with the increase of the width, and the green bar (gray area) declares the penetration depth of edge states around 16 nm (Tab.~\ref{tab:1})}
\end{figure}

We consider QSHE by some calculations of localized band current in a semi-classical viewpoint. Localized band current represents a net current of electrons or holes of each band on each site of the primitive cell. On \textit{n}th site, localized spin-up (spin-down) current is $J_{\uparrow(\downarrow)}(na)=\sum_{k} q v_{\uparrow(\downarrow)}(k)|\psi_{\uparrow(\downarrow)}(k.na)|^{2} $, where $v_{\uparrow(\downarrow)}(k)=\frac{1}{\hbar} \frac{\partial \epsilon_{\uparrow(\downarrow)}(k)}{\partial k} $ is the velocity of spin-up (spin-down) electron or hole and $q$ is the charge of the current-carrier that is located in the energy level of $\epsilon_{\uparrow(\downarrow)}(k)$. Also, $\psi_{\uparrow(\downarrow)}(k.na)$ is the corresponding eigenvector component in \textit{n}th site. Localized band spin current and charge current are $J_{s}(na)=(J_{(\uparrow)}(na)-J_{(\downarrow)}(na))$, and $J_{q}(na)= J_{(\uparrow)}(na)+J_{(\downarrow)}(na)$ respectively. The localized spin-up, spin-down, total spin, and charge currents are calculated for a full band of 100 nm-QWR, and depicted in Fig.~\ref{fig:3}(a). As is evident, for this case, there are two opposite rotational spin-up and spin-down currents along the edge of QWR, named helical current that is exhibited in Fig.~\ref{fig:3}(b). This phenomenon clearly discloses that the charge current is zero regarding the Chern number in the topic of topological insulator $(G_{xy}= G^{\uparrow}_{xy}+G^{\downarrow}_{xy})$, but a topological invariant is defined due to the spin Hall conductance $(G^{(s)}_{xy}= G^{\uparrow}_{xy}-G^{\downarrow}_{xy})$ that is well-known for the quantum spin Hall effect (QSHE).

The repeated pattern of the spin current is observed for the different widths of QWRs as the previous result for 100 nm-QWR. It means QSHE not only exists for wider QWRs, greater than 400 nm in width that we call them bulk samples, but also remains in QWRs with a narrower width around 30 nm, as is depicted in Fig.~\ref{fig:3}(c). As the results declared, the helical spin current exists near the edge of the ribbons, therefore, the maximum value of spin current, for the different widths of QWR, is placed about 3 nm from the edge. Furthermore, we numerically quantify the penetration depth of the localized spin current (LSC). In the Ref~\cite{wada_PhysRevB.83.2011_Bi}, the penetration depth of the edge states is estimated by the solution based on the continuum Hamiltonian that was indicated that the edge states are proportional to $e^{\lambda y}$ $(Re(\lambda) > 0)$. The secular equation determines $l=\lambda^{-1}$ as a penetration depth. In the HgTe/CdTe quantum wells, the penetration depth of the edge states has been calculated by some authors\cite{zhou_PhysRevLett.101.2008_FSQSHE,wada_PhysRevB.83.2011_Bi} to be relatively long, $l\sim 50$ nm. The penetration depth, in our criterion, is determined based on this situation where the LSC is reduced by 90\% of its maximum value. As Table.~\ref{tab:1} shows, the penetration depth, $l$, for QWRs, which are broader than 200 nm, drops slightly. It is near 16~nm for 200~nm, 400~nm, 500~nm, and 800~nm-QWRs. Moreover, the maximum value of the LSC is hovered around 5.21 $\mu A$. So for these samples, the helical spin current, which sustains QSHE, is well localized at the edges, and the bulk is empty. It seems that a different trend happens, when the width of QWRs decreases from 200 nm. In 100 nm and 50 nm-QWRs, $l$ rises steadily to 18 and 19 nm, and the maximum value of the LSC becomes 5.47, 5.27 $\mu A$ respectively. So by decreasing the width of QWRs, the penetration depth and the maximum value of the LSC are increased. An interesting issue occurs when half the width of sample is comparable in size to the penetration depth. In 30 nm-QWR, the penetration depth is about 14.26 nm almost the same to half the width, 15 nm. The helical spin current and the edge states consequently tend to penetrate all the sample widths, extend the entire width of the sample, and dissolve the QSHE. But conversely, the maximum value of the LSC goes upward around 6 $\mu A$ to maintain the QSHE. This indicates there is a permanent contest, on one hand, decreasing the width of QWR attempts to attenuate the helical spin current and annihilate the QSHE. But on the other side, because of the existence of time-reversal modes in BHZ-Hamiltonian, the progressing LSC at the edges tends to support the QSHE. Finally, as Fig.~\ref{fig:3}(c) exhibits the conquering side maintains QSHE by amplifying the LSC at the edges.

\begin{table}[!h]
	\begin{center}
		\caption{\label{tab:1}the results for different widths of QWRs of the prime column values are reported. The second column determines the penetration depth of the edge states in nm; the third column shows an absolute maximum of LSC in $\mu A$ plotted in Fig.~\ref{fig:3}(c); the last column lists distance from the edges, where LSC reaches its maximum value.}
		\begin{tabular}{c c c c}
			\hline \hline
			width of&$l$~(nm)&abs. max.    &position of max.\\
			QWRs~(nm)&       &of LSC~($\mu A$)&of LSC~(nm)\\ \hline
			30      & 14.26   & 5.95  & 3.06         \\ 
			50      & 19.12   & 5.47  & 2.99         \\ 
			100     & 17.88   & 5.27  & 2.94         \\ 
			200     & 16.42   & 5.23  & 2.93         \\ 
			400     & 16.12   & 5.21  & 2.92         \\ 
			500     & 16.08   & 5.21  & 2.92         \\ 
			800     & 16.04   & 5.21  & 2.92         \\ \hline \hline
		\end{tabular}
	\end{center}
\end{table}

\begin{figure}[!h]
	\centering	
	\begin{minipage}[c]{\textwidth}
		\includegraphics[width=1.2\textwidth]{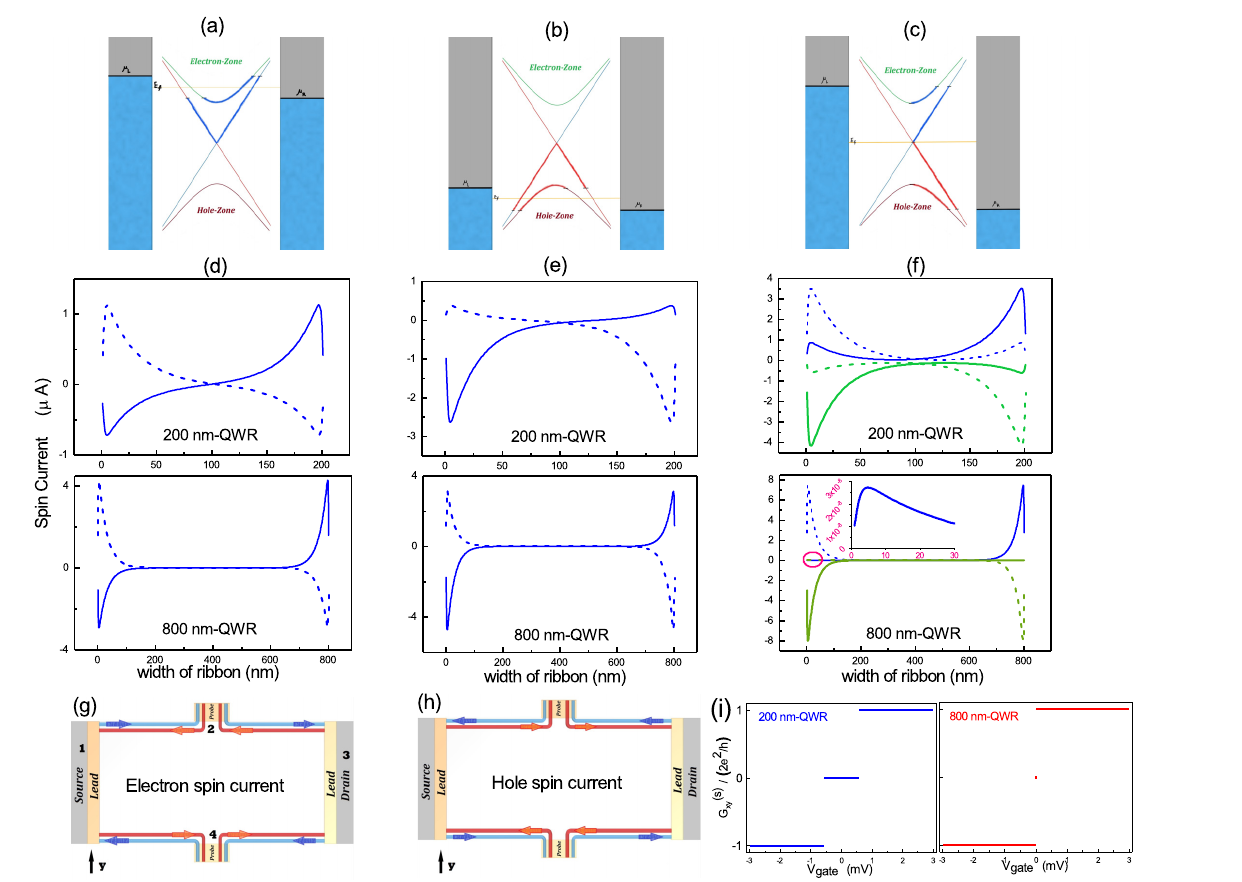}
	\end{minipage}
	\caption{\label{fig:4}(color online). (a,b,c) The schematic views of energy space between source and drain with chemical potentials as $\mu_{L}$ and $\mu_{R}$ respectively are exhibited for 200 nm-QWR (parabolic curve) and 800 nm-QWR (cross lines)  (a) The Fermi energy is moved upper than band gap by gate voltage, so the electron-pockets are current-carrying. (b) The hole-pockets are carriers of current, when we push the Fermi energy toward the hole-zone by applying a negative $v_{gate}$. (c) The Fermi energy is located between the band gap and $\Delta \mu$ is such that one channel of current opens for electrons and another one opens for holes, so both carrier types participate in transmitting current. (d,e,f) The spin-up (solid), spin-down (dashed) are exhibited for 200 nm (top plot) and 800 nm (bottom plot) QWRs (d) electron-pockets, (e) hole-pockets and (f) spin currents for electron carrier (blue, dark) and hole carrier (green, gray) sketched for spin-up (solid) and spin-down (dashed) currents, (inset: shows localized spin-up current in the range up to 30~nm). (g,h) schematic view of two-probes device to investigate spin current direction for (g) electrons and (h) holes for spin-up (blue, lighter) and spin-down (red, darker). (i) The QSH conductance is threefold in the presence of band gap and (left) twofold in the absence of energy gap (right).}
\end{figure}

Right now, we are going to investigate the QSHE by measuring and calculating $G^{(s)}_{xy}$. In this method, the QWR is sandwiched between two leads. Then by applying a bias voltage to spin-up and spin-down electrons and measuring the voltage difference between transverse probes, $G^{\uparrow}_{xy}$, $G^{\downarrow}_{xy}$, and $G^{(s)}_{xy}$ are obtained. Since the filled bands could not pass current through the device, if a gate voltage applies, the Fermi energy can be moved inside the conduction-band region by positive values or inside the valence-band region by a negative one. So it means that electron-pockets in the first case and hole-pockets in the second case are responsible for current-carrying through the device. As mentioned, by employing a gate voltage greater than zero, we can move the Fermi surface to the electron-region and activate electron-pockets to transmit current by energy band. Now, if we apply a bias voltage, the chemical potential equilibrium of left and right junctions is broken that causes electron-pockets to flow by the hash-shaped part of the energy band in Fig.~\ref{fig:4}(a). Spin-up, spin-down, and spin currents corresponding to these carriers are shown in Fig.~\ref{fig:4}(d). As we see in Fig.~\ref{fig:4}(g), spin-up (spin-down) electrons flow clockwise (counter-clockwise) on the edges. By measuring the difference voltage between transverse probes in Fig.~\ref{fig:4}(g) $(\Delta V_{trans}=V_{2}-V_{4}) $, for spin-up electrons $\Delta V_{trans}= \frac{\Delta \mu}{e}$, but for spin-down electrons $\Delta V_{trans}=-\frac{\Delta \mu}{e}$. Using Landauer-Büttiker formalism, in the ballistic regime, we have $I_{x}=\frac{e}{h} \Delta \mu$, thus by using $G_{xy}=\frac{I_{x}}{\Delta V_{trans}}$ we have $G^{\uparrow}_{xy}=\frac{e^{2}}{h}$, $G^{\downarrow}_{xy}=-\frac{e^{2}}{h}$ (the minus sign is due to the opposite direction of spin-down current), and $G^{(s)}_{xy}=\frac{2e^{2}}{h}$ (see~\ref{sec:AppA}). By utilizing a negative gate voltage and entrance to hole-region, all of the above discussion will be to the contrary. The spin-up, spin-down and spin currents are conducted in the opposite direction of the electron-pockets because the holes have a converse spin in energy band (Fig.~\ref{fig:4}(e)), so $G^{\uparrow}_{xy}=-\frac{e^{2}}{h}$, $G^{\downarrow}_{xy}=\frac{e^{2}}{h}$, and $G^{(s)}_{xy}=-\frac{2e^{2}}{h}$. The schematic view of spin-down currents is illustrated in Fig.~\ref{fig:4}(h) for this case. 

In the zero gate voltage, if the voltage difference between the two leads is less than the band gap, there is no energy channel for transfer of charge and therefore the spin Hall conductance will be zero. So QWRs are the ordinary insulator (Fig.~\ref{fig:4}(c)). As the bias voltage increases, because the chemical potential difference between two leads exceeds the band gap, two energy channels open above and below the Fermi energy level. In this situation, electron-pockets and hole-pockets are both responsible for current-carrying. In Fig.~\ref{fig:4}(f), spin-up and spin-down currents are depicted for both types of carriers. As we see, spin-up and spin-down for electrons and holes, do not irrationally flow in the edge of QWRs, so for both of them we have $\Delta V^{\uparrow(\downarrow)}_{trans}=0$, and QSH-conductance vanishes. The detail about the computing of the LSC for bulk samples (say 800 nm) is shown in the inset plot of Fig.~\ref{fig:4}(f). As we see, the LSC behaves like the form of the narrower samples; as a rule of thumb, about $10^5$ electrons with spin-up per second pass along the bottom edge in 800 nm-QWR that represents a considerable amount. This confirms the existence of non-helical LSC in the bulk samples. In Fig.~\ref{fig:4}(i), we exhibit a schematic diagram of QSHE invariant ($G^{(s)}_{xy}=\frac{2e^{2}}{h}$) versus the gate voltage. We find that QWRs experience a topological phase transition by making gradual adjustment of the gate voltage. In zero gate voltage, all of the QWRs are in the ordinary phase, but by applying a gate voltage, when one of the electron-pocket channels or the hole-pocket channels opens for the transmission of charge, QWRs transition to the QSH phase.

\begin{figure}[!h]
	\centering	
	\begin{minipage}[c]{0.29\textwidth}
		\includegraphics[height=3.2cm,width=\textwidth]{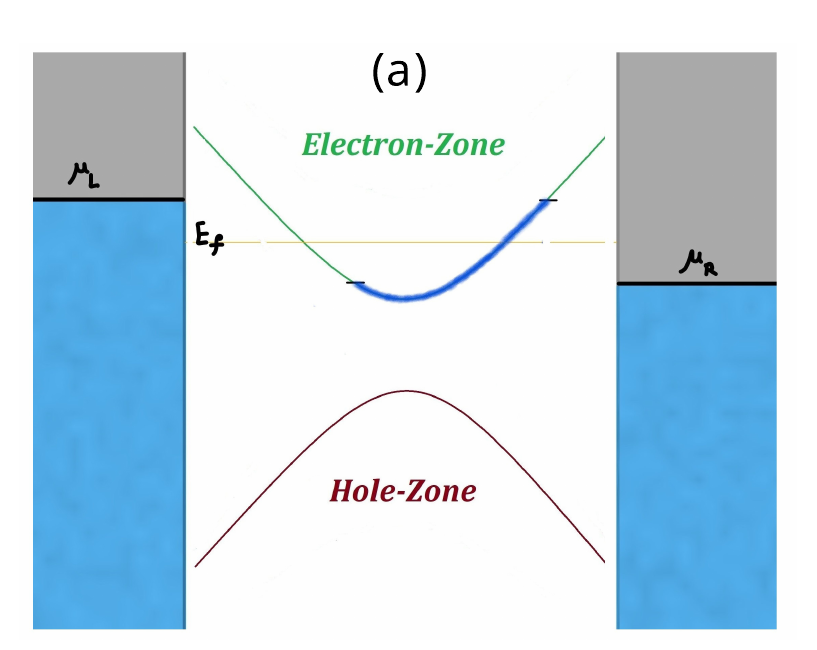}
	\end{minipage}
	\hspace{-\abovedisplayskip}
	\begin{minipage}[c]{0.35\textwidth}
		\includegraphics[width=\textwidth]{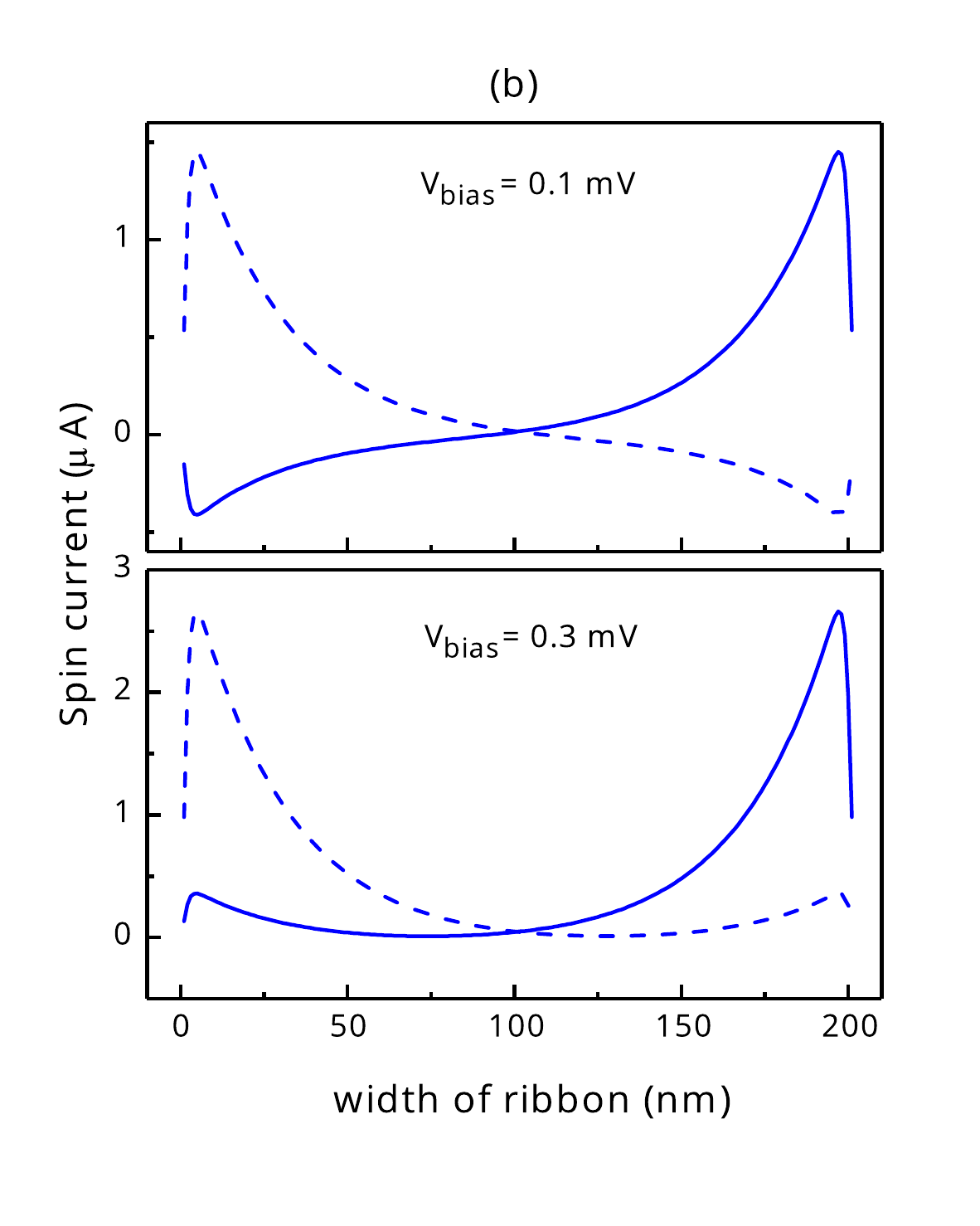}
	\end{minipage}
	
	\vspace{-\abovedisplayskip}	
	\caption{\label{fig:5}(color online).(a) The schematic view of energy space between source and drain with chemical potentials as $\mu_{L}$ and $\mu_{R}$ respectively are shown, the hash shaped part (blue hash-part) indicates the electrons are responsible for spin currents. (b) The spin-up (solid) and spin-down (dashed) currents are depicted for $V_{bias}=\frac{\Delta \mu}{e}=0.1~mV$ (top), and  $V_{bias}=\frac{\Delta \mu}{e}=0.3~mV$ (bottom) for 200 nm-QWR.}
\end{figure}

In more detailed consideration, it is discovered that there is a topological phase transition by adjusting the bias voltage. In Fig.~\ref{fig:5}(b), the spin-up, spin-down, and spin currents are depicted for 200 nm-QWR with the gate voltage $V_{g}=0.5~meV$, when the bias voltage is $V_{bias}=\frac{\Delta \mu}{e}=0.1~mV$ and $ V_{bias}=\frac{\Delta \mu}{e}= 0.3~mV$. We see that in the lower bias voltage there are rotational spin-up and spin-down currents on the edge, so $\Delta V^{\uparrow(\downarrow)}_{trans}=\frac{\Delta \mu}{e} (-\frac{\Delta \mu}{e})$ and QSH topological invariant is non-zero, but for higher bias voltage there are not rotational spin-up and spin-down currents, so $\Delta V^{\uparrow(\downarrow)}_{trans}=0$ and QSH topological invariant becomes zero (see~\ref{sec:AppA}). We deduce the increase of bias voltage disrupts rotational spin-up and spin-down currents and changes the topological phase of QWRs. In our calculations, this phenomenon has been seen for whole QWRs with various width and when the hole-pocket are also current-carriers.
\subsection{\label{sec:ElectricF}QWRs in a transverse Electric Field}

\begin{figure}[!h]
	\centering
	\vspace{-\abovedisplayskip}
	\begin{minipage}[t]{0.5\textwidth}
		\includegraphics[width=\textwidth]{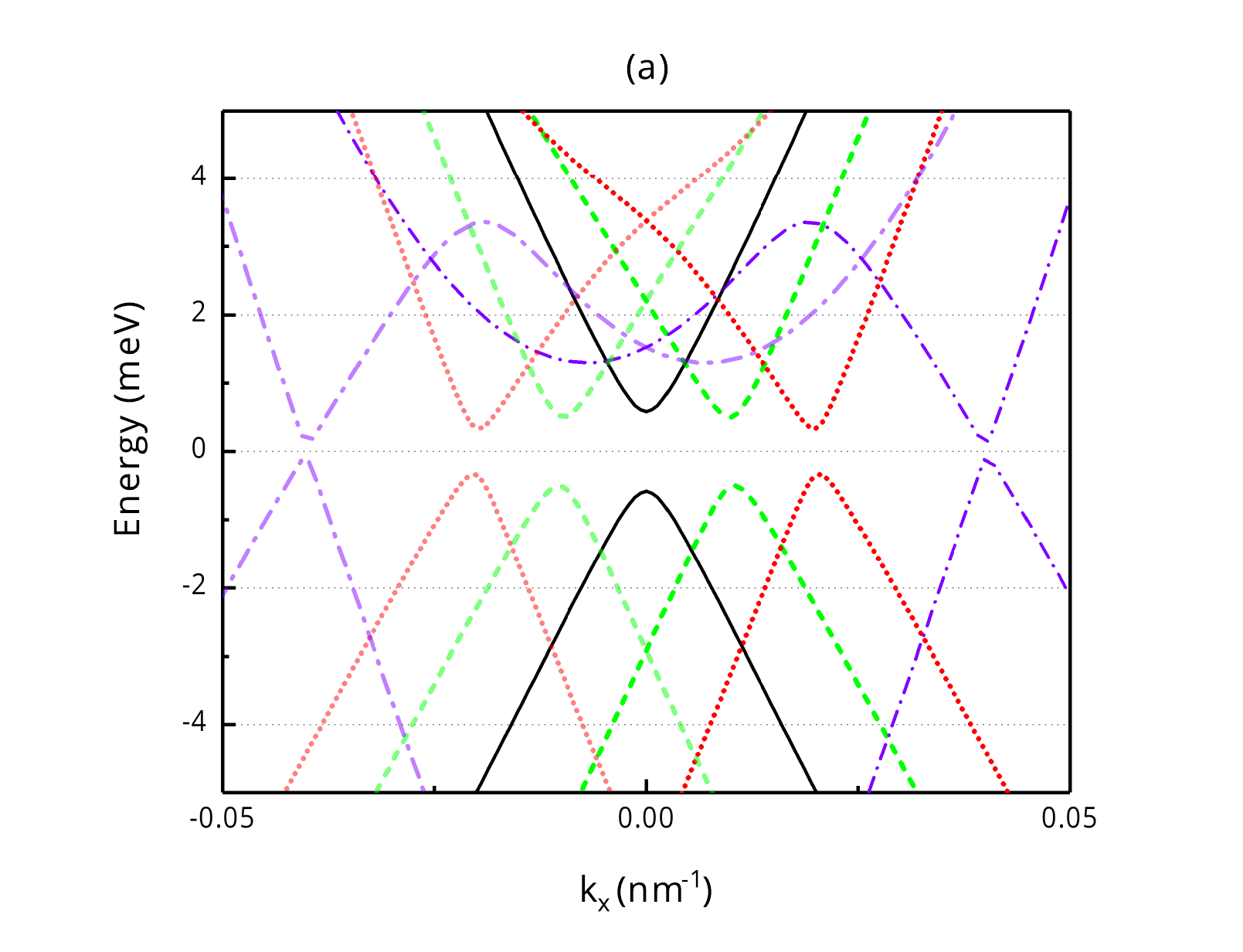}
	\end{minipage}
	\hspace{-\abovedisplayskip}
	\begin{minipage}[t]{0.5\textwidth}
		\includegraphics[width=\textwidth]{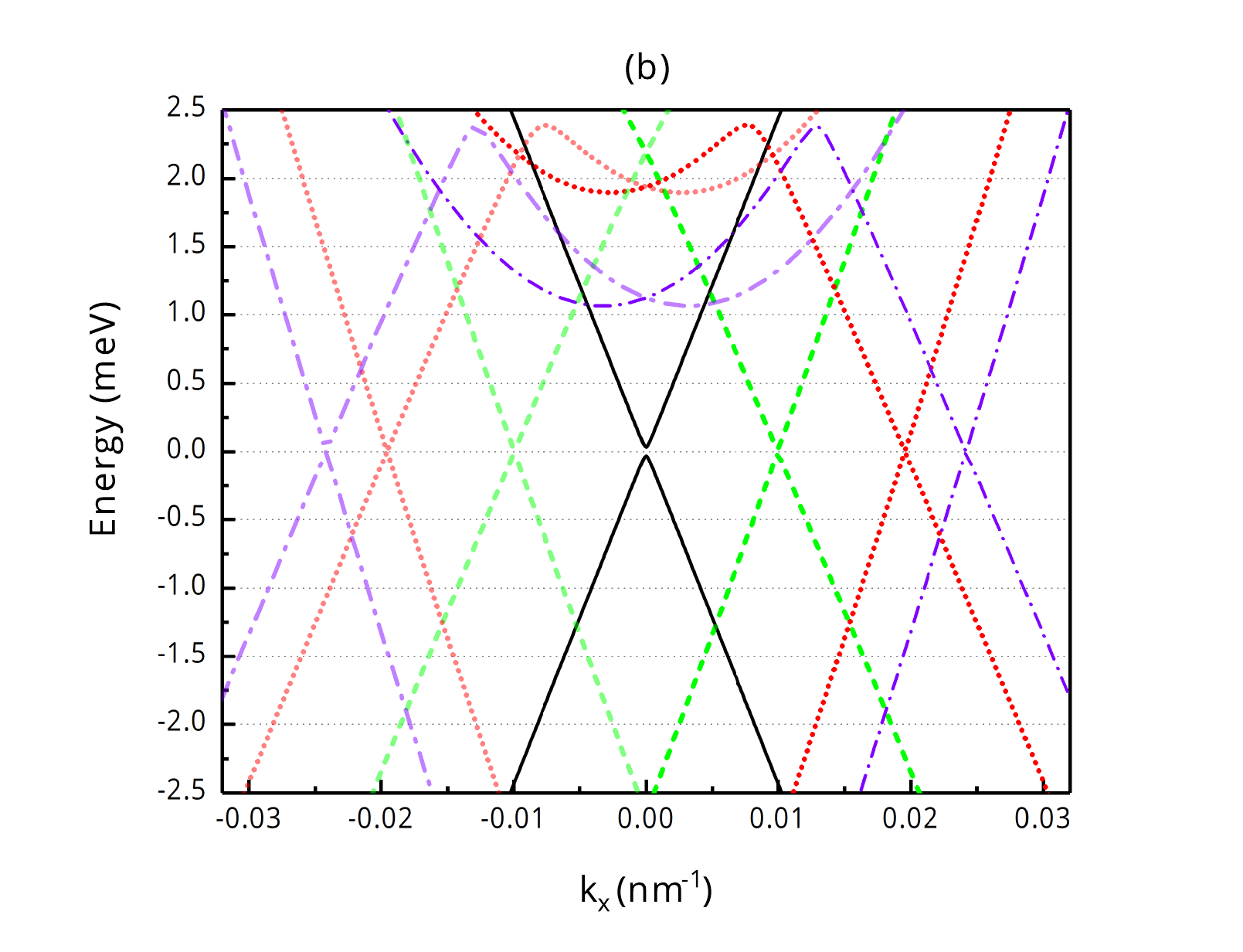}
	\end{minipage}	
	\hspace{-\abovedisplayskip}
	\begin{minipage}[t]{0.5\textwidth}
		\includegraphics[width=\textwidth]{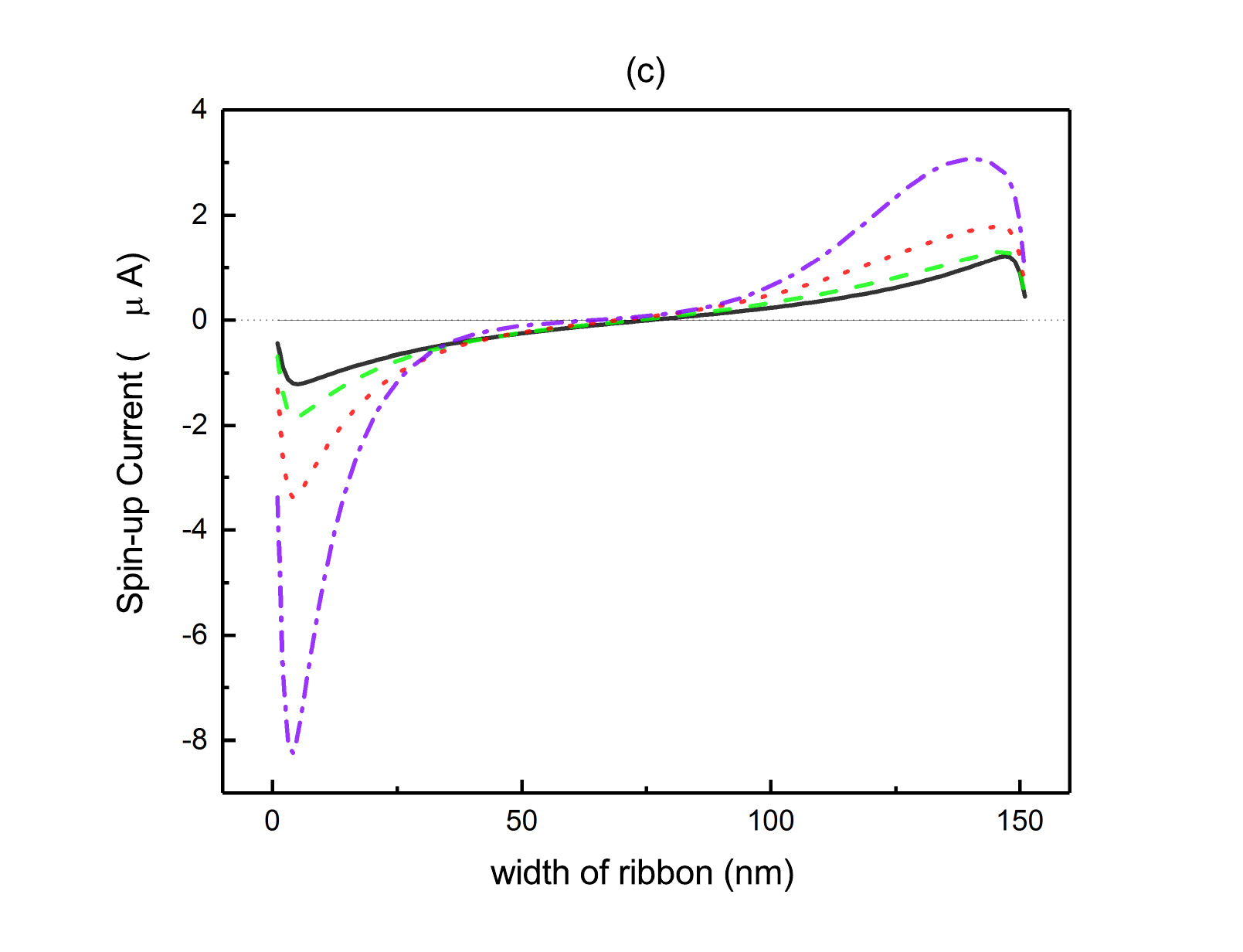}
	\end{minipage}	
	\hspace{-\abovedisplayskip}
	\begin{minipage}[t]{0.5\textwidth}
		\includegraphics[width=\textwidth]{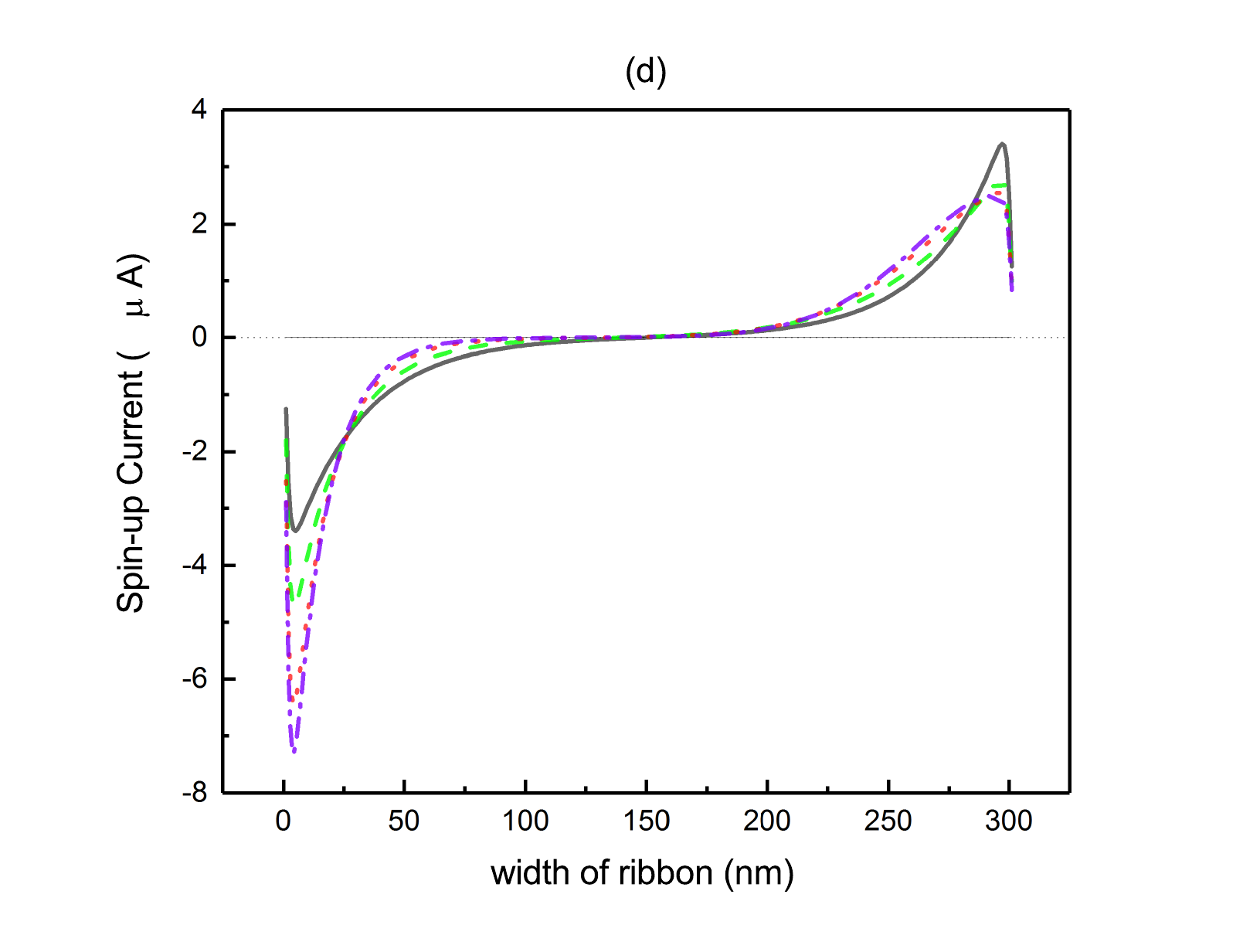}
	\end{minipage}	
	\caption{\label{fig:6} (color online). Band structure is depicted in the presence of an electric field, (a) for 150 nm-QWR, strength of electric field: 0 (black, solid), $0.05 \times 10^{6}~\frac{V}{m}$ (green, dashed),  $0.1 \times 10^{6}~\frac{V}{m}$ (red, dotted), and  $0.2 \times 10^{6}~\frac{V}{m}$ (violet, dashed dot),  (b) for 300 nm-QWR, strength of electric field: 0 (black, solid), $0.02 \times 10^{6}~\frac{V}{m}$ (green, dashed),  $0.04 \times 10^{6}~\frac{V}{m}$ (red, dotted), and  $0.05 \times 10^{6}~\frac{V}{m}$ (violet, dashed dot); The localized spin-up current is shown, (c) for 150 nm-QWR, strength of electric field: 0 (black, solid), $0.05 \times 10^{6} ~\frac{V}{m}$ (green, dashed),  $0.1 \times 10^{6}~\frac{V}{m}$ (red, dotted), and  $0.2 \times 10^{6}~\frac{V}{m}$ (violet, dashed dot). (d) for 300 nm-QWR, strength of electric field: 0 (black, solid), $0.02 \times 10^{6}~\frac{V}{m}$ (green, dashed),  $0.04 \times 10^{6} \frac{V}{m}$ (red, dotted), and  $0.05 \times 10^{6}~\frac{V}{m}$ (violet, dashed dot).}
\end{figure}

The application of a transverse electric field brings significant effects. First, the curve of spin-up and spin-down bands are separated from each other in k-space. This separation between bands is dependent on the intensity of the applied transverse electric field and the width of QWRs (Fig.~\ref{fig:6}(a,b)). The cause of this phenomenon is giving the spin current a push from one edge towards the center of the QWR by an electric field (Fig.~\ref{fig:6}(c,d)), and robustness of time-reversal symmetry. In the absence of an electric field, there is a localized spin-up (spin-down) current with positive (negative) velocity in the upper edge and vice versa in the lower edge (Fig.~\ref{fig:2}(a)). Therefore, the spin-up (spin-down) electrons are in the right (left) hand side of the electronic band with the positive (negative) group velocity. In this state, the distribution of electrons on the edges of the QWR is symmetric, so the band structure and spin current are symmetric too. By applying a transverse electric field, the distribution of electrons becomes asymmetric. Our calculations show that electrons push from the lower edge into the upper one (Fig.~\ref{fig:6}(c,d)). Furthermore, the slope of the left and right-hand side of the electronic band becomes lesser and greater respectively. Besides, the position of band gap shifts towards the positive direction of the momentum axis, k, for spin-up electrons, and this happens in reverse for the spin-down electrons because of time-reversal symmetry. Hence, it seems that by separation of spin-up and spin-down energy bands, the relaxation time of spin increases, and therefore, the QSH regime becomes stronger. In another case, the increase in the electric field leads to a gradual decrease in the band gap. As the Fig.~\ref{fig:6}(a) shows, the band gap of 150 nm-QWR gradually drops 24\% of its value (1.16 meV) by application of the electric field about $0.2 \times 10^{6}~\frac{V}{m}$. Moreover, in Fig.~\ref{fig:6}(b), the band gap of 300 nm-QWR almost eliminates around $0.05 \times 10^{6}~\frac{V}{m}$ of the electric field.

\begin{figure}[!h]
	\centering
	\begin{minipage}[c]{0.5\textwidth}
		\includegraphics[width=\textwidth]{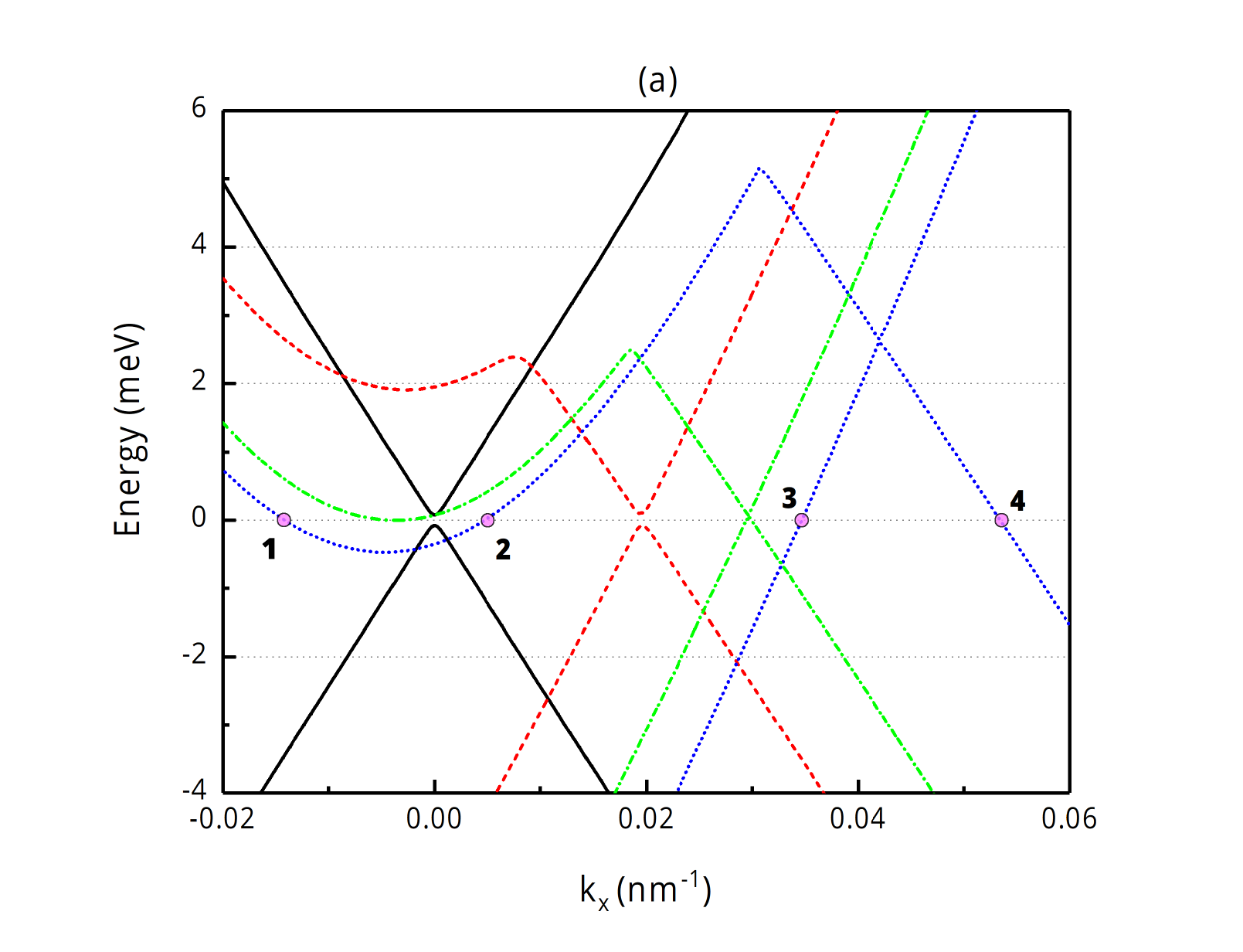}
	\end{minipage}	
	\hspace{-\abovedisplayskip}
	\begin{minipage}[c]{0.5\textwidth}
		\includegraphics[width=\textwidth]{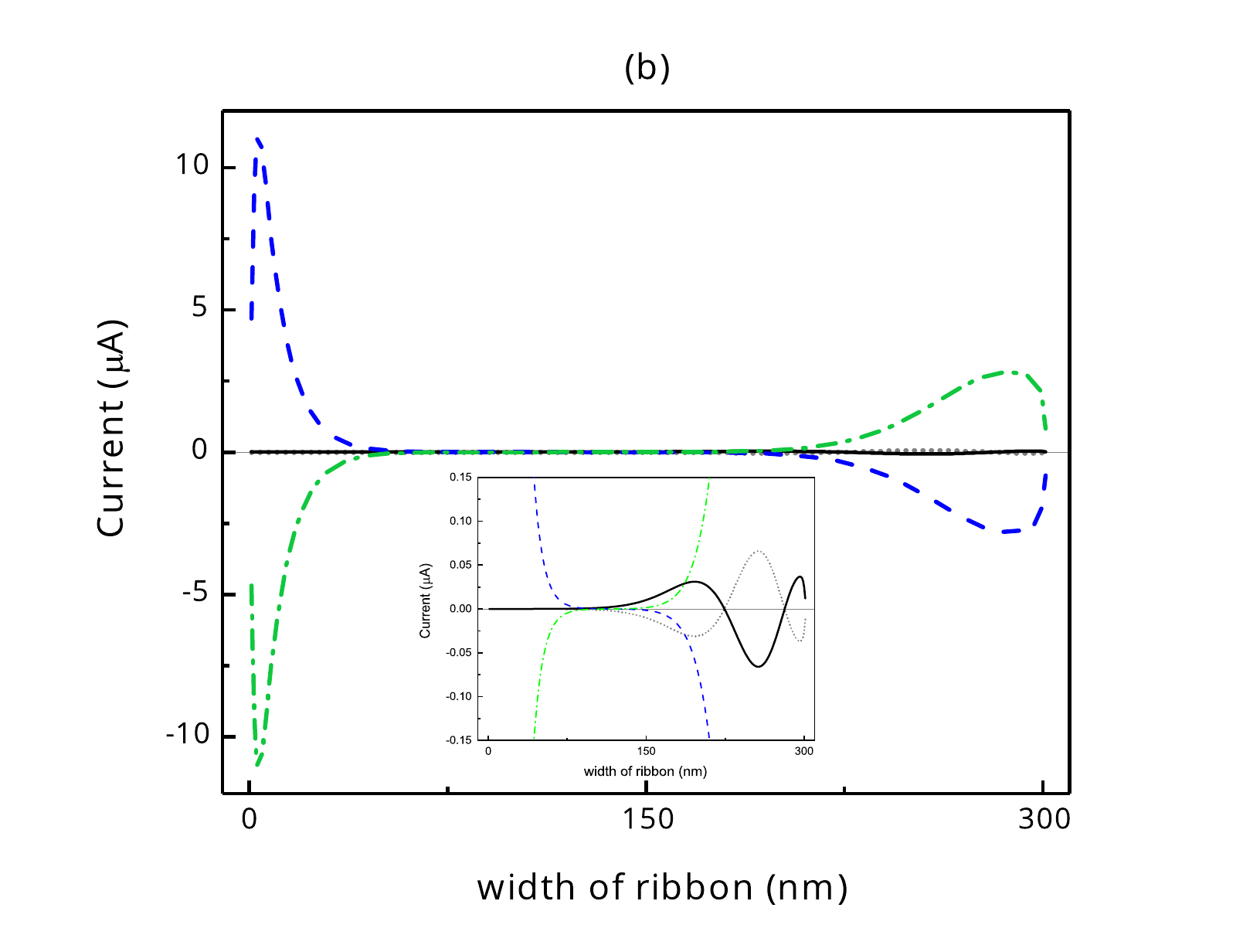}
	\end{minipage}	
	\caption{\label{fig:7} (color online). (a) The band structure of 300 nm-QWR is depicted for the strength of electric field: 0 (black, solid), $0.04 \times 10^{6}~\frac{V}{m}$ (red, dashed), $0.06 \times 10^{6}~\frac{V}{m}$ (wine, dashed dot), and  $0.09 \times 10^{6}~\frac{V}{m}$ (blue, dot). (b) The spin current of the hole-pocket (3 to 4): The spin-up (blue, dashed) and spin-down (green, dashed dot) currents are shown for the electric field with $0.09 \times 10^{6}~\frac{V}{m}$ in strength. (inset): shows an energy window between -0.015~meV to 0.015~meV, the spin current of electron-pocket (1 to 2) is depicted for spin-up (black, solid) and spin-down (gray, dot). }
\end{figure}

In Fig.~\ref{fig:7}(a), the spin-up bands are depicted in various electric field in strength for 300nm-QWR. We see that in the presence of an electric field about $0.06 \times 10^{6}~\frac{V}{m}$ (threshold of the electric field), the electron-like and hole-like bands overlap to each other. So the band gap closes absolutely, and two channels of current (electron and hole channels) open in the zero gate voltage. In this situation, the electron-pocket and hole-pocket current, which is shown in \ref{fig:7}(b), from point 1 to point 2, and point 3 to point 4 respectively, are responsible for the transport of current. The electron-pocket causes a spin current in the middle of the upper half that is not helical and considers no remarkable behavior (the inset plot of Fig.~\ref{fig:7}(b)), but for the hole-pocket, the spin current is helical that shows a topological state with non-zero spin Hall conductance. According to our previous discussion, when the Fermi energy is in the middle of band gap, QWRs are in the trivial state ($G_{xy}^{(s)}=0$), but by increasing the strength of transverse electric field, the band gap decreases, and in the threshold of an electric field, the overlap takes places. Consequently, a helical hole current produces, and a topological phase transition occurs. Table.~\ref{tab:2} reports the threshold of the electric field for various QWRs in width. This indicates the electric field threshold decreases widely by increasing the width of QWRs.

\begin{table}[!h]
	\begin{center}
		\caption{\label{tab:2} This table reports for what values of the electric field in $10^{6}~\frac{V}{m}$ unit, in the second column, the Fermi surface is crossed by QWR energy bands with different widths in the first column.}
		\begin{tabular}{c c|c c} 
			\hline \hline
			Width of     &Electric field			&Width of      &Electric field\\
			QWRs (nm)     &Threshold   		        &QWRs (nm)      &Threshold \\ 
			\hline
			150        & 0.4       		&400	 	&0.038        \\ 
			200        & 0.2        	&500	 	&0.026        \\ 
			250        & 0.08       	&600	 	&0.020        \\ 
			300        & 0.06       	&700	 	&0.016        \\ 
			350        & 0.05       	&800	 	&0.013        \\  
			\hline \hline
		\end{tabular}
	\end{center}
\end{table}

\begin{figure}[!h]
	\centering
	\hspace{-\abovedisplayskip}
	\begin{minipage}[c]{0.49\textwidth}
		\includegraphics[width=\textwidth]{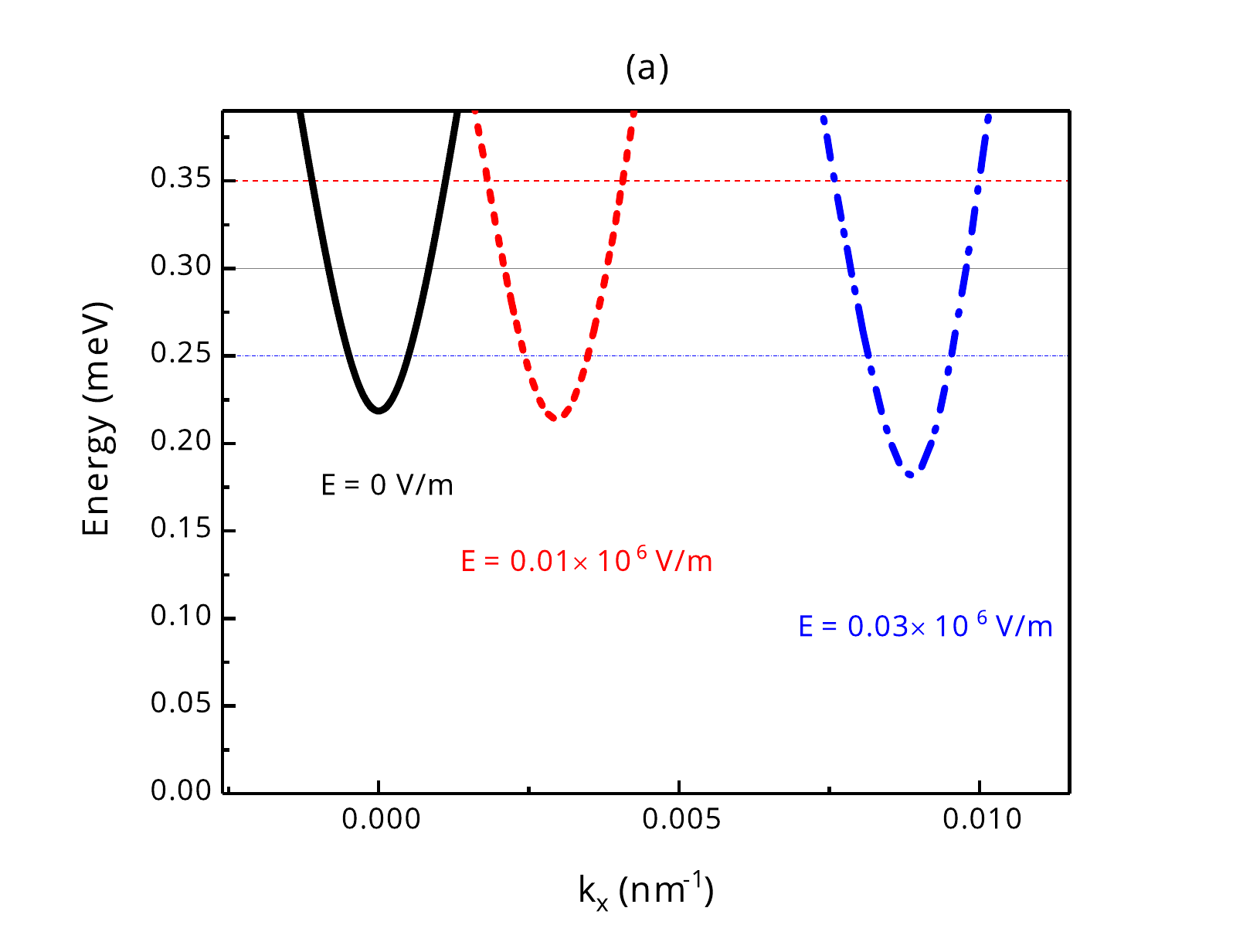}
	\end{minipage}
	\centering	
	\hspace{-\abovedisplayskip}
	\begin{minipage}[c]{0.49\textwidth}
		\includegraphics[width=\textwidth]{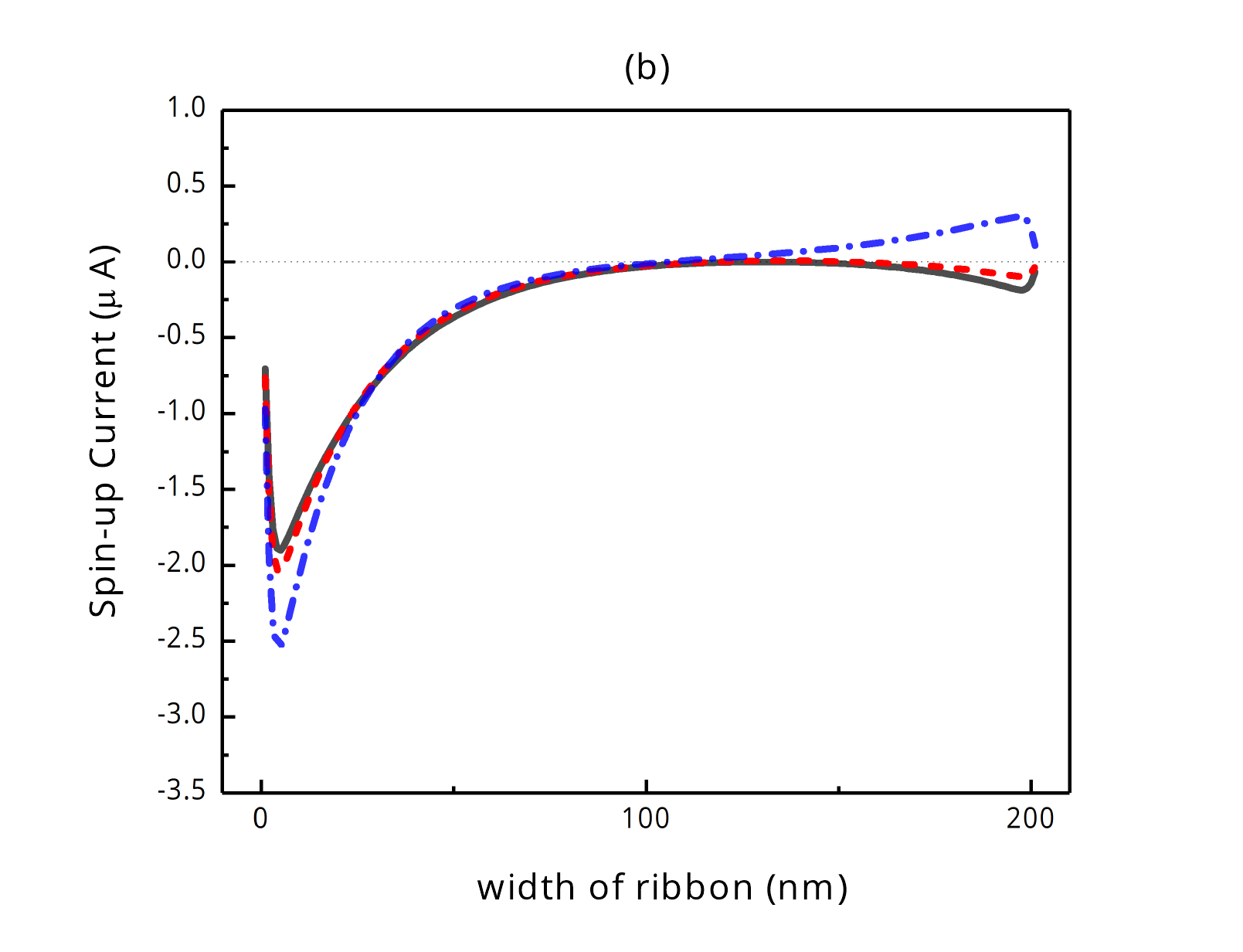}
	\end{minipage}\\%
	
	\caption{\label{fig:8} (color online). (a) The bands and (b) the spin-up current for 200 nm-QWR by employing $v_{gate}=0.3~mV$, and $v_{bias}=\frac{\Delta \mu}{e}=0.1~mV$ in different electric fields: 0 (black, solid), $0.01 \times 10^6~\frac{V}{m}$ (red, dashed), and $0.03 \times 10^6~ \frac{V}{m}$(blue, dashed dot).}
\end{figure}

Our calculations present another evidence that applying a transverse electric field could make the QSHE regime strong in QWRs. It is found for a QWR with a given width, gate and bias voltages, which is in the trivial state, it means that the spin current is not helical. By application of a transverse electric field and gradual increasing, a phase transition happens from trivial to topological that indicates spin current is helical. In Fig.~\ref{fig:8}(a,b), the band and spin current of electron for 200~nm-QWR in gate voltage 0.3 mV and bias voltage 0.1 mV, are depicted in various transverse electric fields. We observe that the helical spin currents no longer exist in zero electric fields, but in the presence of an electric field of magnitude $0.02\times 10^6~\frac{V}{m}$, a helical spin current occurs that shows a topological state. 

\section{\label{sec:Discuss}Experimental Detection and Application}
\begin{figure}[!h]
	\centering
	\hspace{-\abovedisplayskip}
	\includegraphics[width=0.8\textwidth]{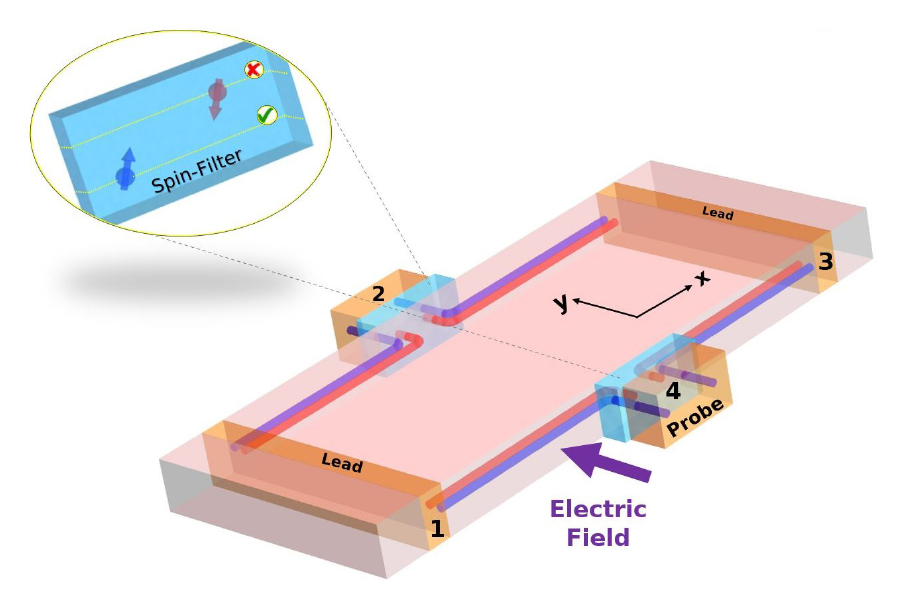}
	
	\caption{\label{fig:9} (color online). Schematic view of a spin-filtered multi-terminal experimental setup; only spin-up electrons exist in voltage probes 2 and 4. A uniform in-plane electric field can be applied in y-direction for tuning topological behavior.}
\end{figure}

Now, we briefly discuss the experimental detection of QSH state in QWR. A full spin-polarized setup can be used to determine the spin Hall conductance $G^{(s)}_{xy}$. Therefore, we want to focus only on purely spin-up or spin-down electrons measurements, similar to schematic device that is shown in Fig.~\ref{fig:9}. As we told in Sec.~\ref{sec:TW}, by using the positive values of $v_{gate}$, the current-carriers are electrons. Then by applying a bias voltage difference between terminal 1 and 3, a helical spin-current would be established (Sec.~\ref{sec:AppA}), and there is a transverse voltage difference between probes 2 and 4. By increasing of bias voltage, in a critical point, the rotational current vanishes and system goes to a non-helical spin-current mode, so transverse voltage difference becomes zero suddenly (as we will calculate in Sec.\ref{sec:AppA}). This critical point, Based on Sec.~\ref{sec:ElectricF}, are tunable by a transverse electric field and the size of ribbons. Thus, it seems that we can use this experimental setup as an electronic switcher in transistors, for example. It can be tuned by gate and bias voltages, and in-plane electric field stregnth.

Producing of full spin-polarized current is much more experimentally challenging. So, in our proposed setup, we can use the voltage probes, which are purely spin filtered devices. It seems that by a weak transmission of current from the voltage probe (probes 2 and 4 in Fig.~\ref{fig:9}), the spin filtered device do its duty well. 

\section{\label{sec:Summary}Summary}
In summary, we studied the QSH regime in a finite-size quantum well of HgTe/CdTe. By calculation of localized spin current along the width of ribbons, and observation of the helical spin current for energy bands in the transport approach, we saw the QSH regime could exist even for narrower QWRs with a considerable band gap. By attaching two leads to the QWR, and calculating the localized spin current and QSH conductance using Landauer-Büttiker formalism, we observed a topological phase transition had occurred by adjustment of the bias voltage. It seemed this transition had takes placed in a critical current, similar to the superconductors. By affecting a transverse electric field, we realized that the spin-up and spin-down bands separated from each other, the band gap decreased and the helical edge current became robust in front of applied bias voltage. Consequently, it seemed the transverse electric field had supported the QSH regime. Although this experiment has not yet performed, it seems these features may pave a direct way for electric control of the edge state transport property.

\section{\label{sec:Thanks}Acknowledgments}
It is a pleasure to thank Dr. S. M. Fazeli for helpful discussions. The calculations were partly performed at the HPC center at University of Qom.

\appendix
\section{\label{sec:AppA}the Landauer-Büttiker formalism for QSHE}

In a multi-terminal device, the left side fills up to the energy level $\mu_{L}$, which is a little more than the right-hand side energy $\mu_{R}$, the conductance is given by a linear response formula, $G=\frac{I}{\Delta V} (-e\Delta V=\mu_{L}-\mu_{R})$. In terms of inter-terminal transmission coefficient, $T_{ij}$ can be seen as the product of number of modes and the transmission probability from \textit{j}th probe to \textit{i}th probe. In the equilibrium condition, we must have $\sum_{j \ne i}^{ }T_{ji}=\sum_{j \ne i}^{ }T_{ij}$, which enables us to write the current in the \textit{i}th terminal as the form $I_{i}=\frac{e^2}{h} \sum_{j \ne i}^{ } T_{ij}(V_{i}-V_{j})$, and the voltage is related with the Fermi energy in the \textit{i}th probe through $\mu_{i}=eV_{i}$, but we can generally write $I_{i}=\frac{e}{h} \sum_{j \ne i}^{ }(T_{ji} \mu_{i}-T_{ij} \mu_{j})$. This formula enables us to write the multi-terminal conductance and resistance in the compact form of matrices. In the QSH system the transmission coefficient with spin-up from one terminal to its neighbor terminal in the clockwise direction is $T^\uparrow_{ij}=1$, and in the counter-clockwise direction is $T^\uparrow_{ji}=0$. This is exactly the opposite for spin-down, means $T^\downarrow_{ij}=0$ and $T^\downarrow_{ji}=1$.

In a four-terminal measurement device, in case~\textit{i}, with the helical edge states, the transmission coefficients for an electron with spin-up are $T^{\uparrow}_{43}=T^{\uparrow}_{32}=T^{\uparrow}_{21}=T^{\uparrow}_{14}=1$, and 0 otherwise, and the transmission coefficients for an electron with spin-down are $T^{\downarrow}_{12}=T^{\downarrow}_{23}=T^{\downarrow}_{34}=T^{\downarrow}_{41}=1$, and 0 otherwise. In case~\textit{ii} that the helical edge states no longer exist similarly, we have $T^{\uparrow(\downarrow)}_{34}=T^{\uparrow(\downarrow)}_{43}=T^{\uparrow(\downarrow)}_{32}=T^{\uparrow(\downarrow)}_{23}=T^{\uparrow(\downarrow)}_{21}=T^{\uparrow(\downarrow)}_{12}=T^{\uparrow(\downarrow)}_{41}=T^{\uparrow(\downarrow)}_{14}=1$, and 0 otherwise. From the Landauer-Büttiker formalism, spin-up currents of the case~\textit{i} and \textit{ii} are respectively,

\begin{equation}
\label{eqn:5}%
\left( {\begin{array}{cccc}
	I^{\uparrow}_{1}  \\
	I^{\uparrow}_{2}  \\
	I^{\uparrow}_{3}  \\
	I^{\uparrow}_{4}  
	\end{array} } \right)_{\textit{i}}
=\frac{e^2}{h}
\left( {\begin{array}{cccc}
	1&0&0&-1  \\
	-1&1&0&0  \\
	0&-1&1&0  \\
	0&0&-1&1  \\
	\end{array} } \right)
\left( {\begin{array}{cccc}
	V_{1}  \\
	V_{2}  \\
	V_{3}  \\
	V_{4}  \\
	\end{array} } \right),
\end{equation}

\begin{equation}
\label{eqn:6}%
\left( {\begin{array}{cccc}
	I^{\uparrow}_{1}  \\
	I^{\uparrow}_{2}  \\
	I^{\uparrow}_{3}  \\
	I^{\uparrow}_{4}  \\
	\end{array} } \right)_{\textit{ii}}
=\frac{e^2}{h}
\left( {\begin{array}{cccc}
	2&-1&0&-1  \\
	-1&2&-1&0  \\
	0&-1&2&-1  \\
	-1&0&-1&2  \\
	\end{array} } \right)
\left( {\begin{array}{cccc}
	V_{1}  \\
	V_{2}  \\
	V_{3}  \\
	V_{4}  \\
	\end{array} } \right),		
\end{equation} 

\noindent and for spin-down current are

\begin{equation}
\label{eqn:7}%
\left( {\begin{array}{cccc}
	I^{\downarrow}_{1}  \\
	I^{\downarrow}_{2}  \\
	I^{\downarrow}_{3}  \\
	I^{\downarrow}_{4}  \\
	\end{array} } \right)_{\textit{i}}
=\frac{e^2}{h}
\left( {\begin{array}{cccc}
	1&-1&0&0  \\
	0&1&-1&0  \\
	0&0&1&-1  \\
	-1&0&0&1  \\
	\end{array} } \right)
\left( {\begin{array}{cccc}
	V_{1}  \\
	V_{2}  \\
	V_{3}  \\
	V_{4}  \\
	\end{array} } \right),
\end{equation}
\noindent 
and for the case~$\textit{ii}$, the T-matrix for spin-down current is equal to spin-up one.

Let set the voltage at terminal 1 and 3 as $V_1=\mu_{L}/e$, and $V_3=\mu_{R}/e$ when the terminals 2 and 4 are voltage probes (see Fig.~\ref{fig:9}). Now, by using a spin filter, let only the spin-up current is established, then apply the constrains for voltage probes, $I_2^\uparrow = I_4^\uparrow = 0$, thus in the case~\textit{i} we can write $I^\uparrow_2=\frac{e^2}{h}(-V_1+V_2)=0$, and $I^\uparrow_4=\frac{e^2}{h}(-V_3+V_4)=0$, so  $V_2=\mu_{L}/e$ and $V_4=\mu_{R}/e$, but in the case~\textit{ii}, $V_2=V_4=\frac{1}{2} (\mu_{L}+\mu_{R})/e$. Therefore, for the spin currents, we observe a completely different state from the case~\textit{i}.
Finally, since the conductance is $G_{xy}^{\uparrow(\downarrow)}=\frac{I_x}{\Delta V_y}=\frac{I^{\uparrow (\downarrow)}_1}{(\pm)(V_2-V_4)}$. we have in the case~\textit{i}, $G^{\uparrow (\downarrow)}_{xy}=\pm1$ in the unit of $\frac{e^2}{h}$, and in the case~\textit{ii}, $G^{\uparrow (\downarrow)}_{xy}=0$.

\section*{References}

\vspace{15pt}
\bibliography{abaradaranbib}

\end{document}